\documentclass[math]{article}
\usepackage{amssymb}
\usepackage{amstex}
\usepackage{amsthm}
\newcommand{\ccc}{{\Bbb C}}
\newcommand{\fff}{{\Bbb F}}

\newcommand{\zzz}{{\Bbb Z}}
\newcommand{\nnn}{{\Bbb N}}
\newcommand{\ffff}{{\frak{F}}}

\newcommand{\ind}{}
\newenvironment{thm}[1]{\smallskip\ind{\sf #1.}\sl}{\smallskip}

\begin{document}
\title{Fusion rules for N=2 superconformal modules}
\maketitle
\begin{center}
Minoru Wakimoto
\end{center}
\begin{center}
{\small
Graduate School of Mathematics, Kyushu University, 
Fukuoka 812-8581, Japan}
\end{center}
\vspace{7mm}
\begin{center}
{\bf Abstract}
\end{center}
\begin{center}
\vspace{3mm}
\parbox{12cm}{
In this note we calculate the fusion coefficients for minimal series 
representations of the N=2 superconformal algebra by using a modified 
Verlinde's formula, and obtain associative and commutative fusion 
algebras with non-negative integral fusion coefficients at each level.}
\end{center}
\vspace{7mm}
\section*{0. Introduction.}
As is known well, fusion algebras are constructed associated to 
each level of affine Lie algebras or W-algebras, and the Verlinde's 
theorem has a crucial importance for explicit calculation of fusion 
coefficients in terms of the matrices of modular transformation $S =
\begin{pmatrix}
0 & -1 \cr
1 & 0
\end{pmatrix}
\in SL(2, \zzz)$.

For superconformal algebras, however, the Verlinde's formula is not 
applicable immediately, since the space spanned by Ramond type 
characters is transformed by $S$ into the space of Neveu-Schwarz 
super-characters although the linear span of Neveu-Schwarz 
characters remains stable under the action of $S$.

Recently P. Minces, M. A. Namazie and C. N\'u$\tilde{\text{n}}$ez 
\cite{MNN} proposed a kind of modification of the Verlinde's formula 
to obtain fusion coefficients for N=1 superconformal modules.  In this 
note, we show that a similar modification works also for the N=2 
superconformal algebra, giving associative and commutative fusion 
algebras with non-negative integral fusion coefficients at each level.

The author is grateful to Dr. K. Iohara for information on 
the paper \cite{MNN}.

After this work is accomplished, the author is informed from Prof. 
M. Walton about the work \cite{EH} of W. Eholzer and R. H\"ubel, 
in which a generalization of Verlinde's formula and its application 
to the fermionic $W(2, \delta)$-algebras are discussed.

\section*{1. Minimal series representations and transformation matrices 
of N=2 superconformal algebra.}

The modular transformation of characters and super-characters of minimal 
series representations of the N=2 superconformal algebra was obtained 
in \cite{RY} and its proof using
the Ramanujan's mock theta identity was provided in \cite{KW1}. 
In this section, we recall characters of minimal series representations 
of the N=2 superconformal algebra and the formula of their modular 
transformation, following closely the paper \cite{KW1}.

First we consider the theta function
$$f(\tau, z) := e^{\frac{\pi i \tau}{4}}e^{-\pi i z}
\prod^{\infty}_{n=1}
(1-q^n)(1-e^{2 \pi i z}q^{n-1})(1-e^{-2 \pi i z}q^n)
$$
defined for $\tau \in \ccc_+ := \{\tau \in \ccc \, ; \,\ \text{Im}\tau
 >  0 \}$ and $z \in \ccc$, where $q=e^{2 \pi i \tau}$ as usual, and  
Dedekind's $\eta$-function
$$\eta(\tau) := q^{\frac{1}{24}}\prod^{\infty}_{n=1}(1-q^n).$$
We introduce functions
$\varphi^{(\varepsilon)}(\tau, z)$ and $\psi^{(\varepsilon)}(\tau, z)$ 
($\varepsilon = 0 \,\ \text{or} \,\ \frac{1}{2}$) as follows :
\begin{eqnarray*}
\varphi^{(\varepsilon)}(\tau, z) &=& 
\begin{cases}
\frac{\eta(\tau)^3}{f(\tau, z)} &\qquad (\text{if} \,\ \varepsilon =0) \\
q^{-\frac{1}{8}}e^{-\pi i z}\frac{\eta(\tau)^3}{f(\tau, z+ \frac{\tau}{2})} 
&\qquad (\text{if} \,\ \varepsilon =\frac{1}{2}), 
\end{cases}  \cr \cr
\psi^{(\varepsilon)}(\tau, z) &=& 
\begin{cases}
-i \varphi^{(0)}(\tau, z+ \frac{1}{2})
&\qquad (\text{if} \,\ \varepsilon =0) \\
\varphi^{(\frac{1}{2})}(\tau, z+ \frac{1}{2})
&\qquad (\text{if} \,\ \varepsilon =\frac{1}{2}).
\end{cases}
\end{eqnarray*}
We put
\begin{eqnarray}
F(\tau, z_1, z_2) &:=& \frac{\eta(\tau)^3 f(\tau, z_1+z_2)}
{f(\tau, z_1)f(\tau, z_2)} \nonumber \\ \nonumber \\
&=& 
\sum_{j,k \ge 0}u^jv^kq^{jk} -
\sum_{j,k < 0}u^jv^kq^{jk}
\label{eqn:mock}
\end{eqnarray}
\noindent
where $u := e^{2 \pi i z_1}$ and  $v := e^{2 \pi i z_2}$.  The second 
equality in \eqref{eqn:mock} is the mock theta identity (cf. \cite{B} 
and \cite{H}).  For $m \in \nnn$ and $j, k \in \frac{1}{2}\zzz$, we put
\begin{eqnarray*}
F^{(m)}_{j,k}(\tau, z) &=&
q^{\frac{jk}{m}}e^{\frac{2 \pi i (j-k)z}{m}}
F(m\tau, \,\ -z +j \tau, \,\ z+k\tau ), \cr
G^{(m)}_{j,k}(\tau, z) &=&
q^{\frac{jk}{m}}e^{\frac{2 \pi i (j-k)z}{m}}
F\left(m\tau, \,\ -z +j \tau - \frac{1}{2}, \,\ z+k\tau +\frac{1}{2}\right).
\end{eqnarray*}

Given a positive integer $m$, minimal series representations of 
level $m$ of the N=2 superconformal algebra are parametrized by the set 
$A^{(m)} := A^{(m; \frac{1}{2})} \cup A^{(m; 0)}$, where 
{\allowdisplaybreaks %
\begin{align*}
&A^{(m; \frac{1}{2})} := NS^{(m)} := 
\left\{ (j, k) \in \left(\frac{1}{2} + \zzz \right)^2 
\,\ ; \,\ 0 \, < \, j, \, k, \, j+k \, < m \right\} \\
&A^{(m;0)} := R^{(m)} := 
\left\{ (j, k) \in  \zzz^2 \,\ ; \,\ 0 \, < \, j, \, j+k \, < \, m \,\ 
\text{and} \,\ 0 \, \le k \, < \, m \right\},
\end{align*}
}
\noindent
and the character and super-character of the representation corresponding 
to $(j,k) \in A^{(m; \varepsilon)}$ are given as follows : 
\begin{eqnarray*}
\text{ch}^{(m; \varepsilon)}_{j,k} (\tau, z) &=&
\frac{G^{(m)}_{j,k}(\tau, z)}{\psi^{(\varepsilon)}(\tau, z)} \cr
\text{sch}^{(m; \varepsilon)}_{j,k} (\tau, z) &=&
\frac{F^{(m)}_{j,k}(\tau, z)}{\varphi^{(\varepsilon)}(\tau, z)}.
\end{eqnarray*}

We define the action of $S = 
\begin{pmatrix}
0 & -1 \cr
1 & 0
\end{pmatrix}
\in SL(2, \zzz)$
on the linear span of characters and super-characters of level $m$ by
$$\chi \vert_S (\tau, z) :=
e^{-\frac{i \pi z^2}{\tau}(1- \frac{2}{m})}
\chi\left(-\frac{1}{\tau}, \frac{z}{\tau}\right),$$
and put 
\begin{eqnarray}
S^{(m)}_{(j,k),(r,s)} := \frac{2}{m}
e^{\frac{i \pi (j-k)(r-s)}{m}}
\text{sin}\frac{\pi (j+k)(r+s)}{m}
\label{eqn:trans}
\end{eqnarray}
for $(j,k)$ and $(r,s) \in A^{(m)}$.
Then the transformation of (super-)characters under $S$ is given 
in \cite{KW1} as follows :

\begin{thm}{Proposition 1.1}
\begin{eqnarray*}
{\rm ch}^{(m; \varepsilon)}_{j,k}\vert_S &=&
\begin{cases}
\sum\limits_{(j,k) \in NS^{(m)}} S^{(m)}_{(j,k),(r,s)}
{\rm ch}^{(m; \frac{1}{2})}_{r,s} &\qquad {\rm if} \,\ \varepsilon = 
\frac{1}{2} \\
\sum\limits_{(j,k) \in NS^{(m)}} S^{(m)}_{(j,k),(r,s)}
{\rm sch}^{(m; \frac{1}{2})}_{r,s} &\qquad {\rm if} \,\ \varepsilon = 0,
\end{cases} \cr
{\rm sch}^{(m; \varepsilon)}_{j,k}\vert_S &=&
\begin{cases}
\sum\limits_{(j,k) \in R^{(m)}} S^{(m)}_{(j,k),(r,s)}
{\rm ch}^{(m; 0)}_{r,s} &\qquad {\rm if} \,\ \varepsilon = \frac{1}{2} \\
\sum\limits_{(j,k) \in R^{(m)}} S^{(m)}_{(j,k),(r,s)}
{\rm sch}^{(m; 0)}_{r,s} &\qquad {\rm if} \,\ \varepsilon = 0.
\end{cases} 
\end{eqnarray*}
\end{thm}

It is easy to see that
\begin{eqnarray}
\overline{S^{(m)}_{u,v}} = S^{(m)}_{u^{\ast}, v}
\label{eqn:conjugate}
\end{eqnarray}
for all $u, v \in A^{(m)}$, where \lq \lq  $\ast$ "  is the 
{\it conjugation} of $A^{(m; \varepsilon)}$ defined by
\begin{eqnarray*}
(j,k)^{\ast} := 
\begin{cases} 
(k,j) & \qquad \qquad \text{if} \,\ k \ne 0 \\
(m-j,k) & \qquad \qquad \text{if} \,\ k = 0. 
\end{cases}
\end{eqnarray*}

We note that the modular transformation $S$ satisfies
$$\left({\rm ch}^{(m; \varepsilon)}_{u}\vert_S \right)\vert_S = 
{\rm ch}^{(m; \varepsilon)}_{u^{\ast}}$$
for all $u \in A^{(m; \varepsilon)}$, namely
\begin{eqnarray}
\sum_{w \in NS^{(m)}}S^{(m)}_{u,w}S^{(m)}_{w,v}=\delta_{u, v^{\ast}}
\label{eqn:unitary}
\end{eqnarray}
when both $u$ and $v$ belong to $NS^{(m)}$ or both to $R^{(m)}$.

\section*{2. Modified Verlinde's formula and fusion coefficients.}

In this section we fix a positive integer $m$, and calculate the fusion 
algebra of level $m$.  Let us consider the set
$$\fff^{(m)} :=
\bigcup_{\varepsilon + \varepsilon^{\prime} + \varepsilon^{\prime \prime}
= \frac{1}{2} \, \text{or} \, \frac{3}{2}}
\left(A^{(m; \varepsilon)} \times A^{(m; \varepsilon^{\prime})} \times 
A^{(m; \varepsilon^{\prime \prime})}\right).$$
For $(u, u^{\prime}, u^{\prime \prime}) \in \fff^{(m)}$, we define
the {\it fusion coefficients} $N_{u, u^{\prime}, u^{\prime \prime}}$  and
$N_{u, u^{\prime}}^{u^{\prime \prime}}$  as follows :
{\allowdisplaybreaks %
\begin{eqnarray}
N_{u, u^{\prime}, u^{\prime \prime}}
&:=& \sum_{v \in NS^{(m)}}
\frac{S^{(m)}_{u, v} 
S^{(m)}_{u^{\prime}, v} S^{(m)}_{u^{\prime \prime}, v}} 
{S^{(m)}_{(1/2, 1/2), v}}, 
\label{eqn:fusion} \\
N_{u, u^{\prime}}^{u^{\prime \prime}}
&:=& \sum_{v \in NS^{(m)}}
\frac{S^{(m)}_{u, v} 
S^{(m)}_{u^{\prime}, v} \overline{S^{(m)}_{u^{\prime \prime}, v}}} 
{S^{(m)}_{(1/2, 1/2), v}}.
\label{eqn:cofusion}
\end{eqnarray}
}
\noindent
By \eqref{eqn:conjugate}, one has $N_{u, u^{\prime}}^{u^{\prime \prime}} 
= N_{u, u^{\prime}, u^{\prime \prime \ast}}$.
In this section, we are going to prove the following :

\begin{thm}{Theorem 2.1}
Let $(u, u^{\prime}, u^{\prime \prime}) \in \fff^{(m)}$. Then
$N_{u, u^{\prime}, u^{\prime \prime}}$ is equal to $1$ if 
$u=(j,k)$, $u^{\prime}=(j^{\prime}, k^{\prime})$ and 
$u^{\prime \prime}=(j^{\prime \prime}, k^{\prime \prime})$  satisfy 
either one of the following conditions {\rm (F1)} and {\rm (F2)}, 
and $0$ otherwise :
\begin{eqnarray*}
{\rm (F1)} & & \quad (j+j^{\prime}+j^{\prime \prime})-
(k+k^{\prime}+k^{\prime \prime}) =0, \cr
& & \quad \vert (j^{\prime}+k^{\prime})-(j^{\prime \prime}
+k^{\prime \prime})
\vert < j+k <
(j^{\prime}+k^{\prime})+(j^{\prime \prime}+k^{\prime \prime}),  \cr 
& & \! \! \! \! \! {\rm and} \cr
& & \quad (j+k)+(j^{\prime}+k^{\prime})
+(j^{\prime \prime}+k^{\prime \prime})<2m,  \cr \cr
{\rm (F2)} & & \quad (j+j^{\prime}+j^{\prime \prime})-
(k+k^{\prime}+k^{\prime \prime}) = \pm m \cr
& & \! \! \! \! \! {\rm and} \cr
& &\quad \vert (m-j^{\prime}-k^{\prime})-(m-j^{\prime \prime}
-k^{\prime \prime})
\vert < m-j-k \cr & & \qquad \qquad <
(m-j^{\prime}-k^{\prime})+(m-j^{\prime \prime}-k^{\prime \prime}).
\end{eqnarray*}
\end{thm}

To prove this theorem, we consider the following summations :
\begin{eqnarray*}
J(a,b) &:=& \sum_{\Sb j, k \in \zzz \,\ {\rm s.t.} \\
0 < k < m, \\
-k < j < k, \\
j+k \equiv 1 \, {\rm mod} \, 2 \endSb}
e^{iaj}e^{ibk}, \cr
I(a,b) &:=& J(a,b) + J(a, -b).
\end{eqnarray*}
for a positive integer $m$ and complex numbers $a$ and $b$.
The following lemma is an easy consequence from the summation formula of 
a geometric series :

\begin{thm}{Lemma 2.1}
\begin{enumerate}
\item[{\rm 1)}]  If $e^{ia}=e^{ib}=1$, \qquad $J(a,b)= \dfrac{m(m-1)}{2}$.
\item[{\rm 1')}]  If $e^{ia}=e^{ib}=-1$, \qquad $J(a,b)= -\dfrac{m(m-1)}{2}$.
\item[{\rm 2)}]  If $e^{ia}=1$ and $e^{ib} \ne 1$, 
$$J(a,b)= (1-m)\frac{e^{ibm}}{1-e^{ib}}
+\frac{e^{ib}-e^{ibm}}{(1-e^{ib})^2}.$$
\item[{\rm 2')}]  If $e^{ia}=-1$ and $e^{ib} \ne -1$, 
$$J(a,b)= (-1)^m(m-1)\frac{e^{ibm}}{1+e^{ib}}
+\frac{e^{ib}+(-1)^me^{ibm}}{(1+e^{ib})^2}.$$
\item[{\rm 3)}] In case when $e^{ia} \ne \pm 1$,
\begin{enumerate}
\item[{\rm (i)}] if $e^{i(a+b)} = 1$ and $e^{i(a-b)} \ne 1$,
$$J(a,b) = \frac{1}{e^{ia}-e^{-ia}}\left\{m-1-
\frac{e^{i(b-a)}-e^{i(b-a)m}}{1-e^{i(b-a)}}\right\},$$
\item[]
\item[{\rm (ii)}] if $e^{i(a+b)} \ne 1$ and $e^{i(a-b)} = 1$,
$$J(a,b) = \frac{1}{e^{ia}-e^{-ia}}\left\{
\frac{e^{i(a+b)}-e^{i(a+b)m}}{1-e^{i(a+b)}}+1-m \right\},$$
\item[]
\item[{\rm (iii)}] if $e^{i(a+b)} \ne 1$ and $e^{i(a-b)} \ne 1$,
$$J(a,b) = \frac{1}{e^{ia}-e^{-ia}}\left\{
\frac{e^{i(a+b)}-e^{i(a+b)m}}{1-e^{i(a+b)}}-
\frac{e^{i(b-a)}-e^{i(b-a)m}}{1-e^{i(b-a)}}\right\}.$$
\end{enumerate}
\end{enumerate}
\end{thm}

From this, one has the following :

\begin{thm}{Lemma 2.2}
Let $m \in \nnn$ and $a, b \in \frac{\pi}{m}\zzz$. Then
\begin{enumerate}
\item[{\rm 1)}]  if $e^{ia}=e^{ib}=1$, \qquad $I(a,b)= m(m-1)$,
\item[{\rm 1')}]  if $e^{ia}=e^{ib}=-1$, \qquad $I(a,b)= -m(m-1)$,
\item[{\rm 2)}]  if $e^{ia}=1$ and $e^{ib} \ne 1$, 
$$I(a,b)= 
\begin{cases}
-m & \qquad (\text{if} \,\ e^{ibm}=1), \\
m-1+\left(\frac{1+e^{ib}}{1-e^{ib}}\right)^2
& \qquad (\text{if} \,\ e^{ibm}=-1), 
\end{cases}
$$
\item[{\rm 2')}]  if $e^{ia}=-1$ and $e^{ib} \ne -1$, 
$$I(a,b)= 
\begin{cases}
m & \qquad (\text{if} \,\ e^{ibm}=(-1)^m), \\
1-m-\left(\frac{1-e^{ib}}{1+e^{ib}}\right)^2
& \qquad (\text{if} \,\ e^{ibm}=-(-1)^m).
\end{cases}
$$
\item[{\rm 3)}] In case when $e^{ia} \ne \pm 1$,
\begin{enumerate}
\item[{\rm (i)}] if $e^{i(a+b)} = 1$ and $e^{i(a-b)} \ne 1$,
$$I(a,b) = \frac{1-e^{i(a-b)m}}{e^{ia}-e^{-ia}} \cdot
\frac{1+e^{i(a-b)}}{1-e^{i(a-b)}},$$
\item[]
\item[{\rm (ii)}] if $e^{i(a+b)} \ne 1$ and $e^{i(a-b)} = 1$,
$$I(a,b) = \frac{1-e^{i(a+b)m}}{e^{ia}-e^{-ia}} \cdot
\frac{1+e^{i(a+b)}}{1-e^{i(a+b)}},$$
\item[]
\item[{\rm (iii)}] if $e^{i(a+b)} \ne 1$ and $e^{i(a-b)} \ne 1$,
$$I(a,b) = \frac{1-e^{i(a+b)m}}{e^{ia}-e^{-ia}}\left\{
\frac{1+e^{i(a+b)}}{1-e^{i(a+b)}}+
\frac{1+e^{i(a-b)}}{1-e^{i(a-b)}}\right\}.$$
\end{enumerate}
\end{enumerate}
\end{thm}

Now we compute 
\begin{eqnarray}
N_{u, u^{\prime}, u^{\prime \prime}}
&=&N_{(j,k), (j^{\prime}, k^{\prime}), 
(j^{\prime \prime}, k^{\prime \prime})}  \nonumber \\
&=& \sum_{(r,s) \in NS^{(m)}}
\frac{S^{(m)}_{(j,k), (r,s)} 
S^{(m)}_{(j^{\prime},k^{\prime}), (r,s)} 
S^{(m)}_{(j^{\prime \prime}, k^{\prime \prime}), (r,s)}}
{S^{(m)}_{(1/2, 1/2), (r,s)}}.
\label{eqn:summand1}
\end{eqnarray}

Since 
\begin{eqnarray*}
\frac{S^{(m)}_{(j,k), (r,s)}}{S^{(m)}_{(1/2, 1/2), (r,s)}}
&=& e^{\frac{i \pi (j-k)(r-s)}{m}}
\frac{\text{sin}\frac{\pi (j+k)(r+s)}{m}}{\text{sin}\frac{\pi (r+s)}{m}} \cr
&=& e^{\frac{i \pi (j-k)(r-s)}{m}}
\sum^{j+k-1}_{t=0}e^{\frac{i \pi}{m}(r+s)(j+k-1-2t)} \cr
&=& e^{\frac{i \pi (j-k)(r-s)}{m}}
\sum^{j+k-1}_{t=0}e^{\frac{i \pi}{m}(r+s)(-j-k+1+2t)},
\end{eqnarray*}
each summand in the right-hand side of \eqref{eqn:summand1} 
is written as follows :
{\allowdisplaybreaks %
\begin{eqnarray}
& &\frac{S^{(m)}_{(j,k), (r,s)} 
S^{(m)}_{(j^{\prime},k^{\prime}), (r,s)} 
S^{(m)}_{(j^{\prime \prime},k^{\prime \prime}), (r,s)}}
{S^{(m)}_{(1/2, 1/2), (r,s)}}   \nonumber \\
&=&-\frac{1}{2m^2}e^{\frac{i \pi}{m}(
(j+j^{\prime}+j^{\prime \prime})-
(k+k^{\prime}+k^{\prime \prime}))(r-s)} \times \nonumber \\
& &\! \! \! \! \! \! \! \! \! \! \! \! \bigg\{
\sum^{j+k-1}_{t=0} \! \! e^{\frac{i \pi}{m}
(j+k+j^{\prime} \! +k^{\prime} \! +j^{\prime \prime} \! 
+k^{\prime \prime} \! -1-2t)(r+s)} 
\! \! +  \! \! \! \!
\sum^{j+k-1}_{t=0} \! \! e^{\frac{i \pi}{m}
(-j-k+j^{\prime} \! +k^{\prime} \! +j^{\prime \prime} \! 
+k^{\prime \prime} \! +1+2t)(r+s)} 
\nonumber \\
& & \! \! \! \! \! \! \! \! \! \! \! \! +
\sum^{j+k-1}_{t=0} \! \! e^{\frac{i \pi}{m}
(j+k-j^{\prime} \! -k^{\prime} \! -j^{\prime \prime} \! 
-k^{\prime \prime} \! -1-2t)(r+s)}
\! \! +  \! \! \! \!
\sum^{j+k-1}_{t=0} \! \! e^{\frac{i \pi}{m}
(-j-k-j^{\prime} \! -k^{\prime} \! -j^{\prime \prime} \!
-k^{\prime \prime} \! +1+2t)(r+s)} 
\nonumber \\
& & \! \! \! \! \! \! \! \! \! \! \! \! -
\sum^{j+k-1}_{t=0} \! \! e^{\frac{i \pi}{m}
(j+k+j^{\prime} \! +k^{\prime} \! -j^{\prime \prime} \! 
-k^{\prime \prime} \! -1-2t)(r+s)}
\! \! -  \! \! \! \!
\sum^{j+k-1}_{t=0} \! \! e^{\frac{i \pi}{m}
(-j-k+j^{\prime} \! +k^{\prime} \! -j^{\prime \prime} \! 
-k^{\prime \prime} \! +1+2t)(r+s)} 
\nonumber \\
& & \! \! \! \! \! \! \! \! \! \! \! \! -
\sum^{j+k-1}_{t=0} \! \! e^{\frac{i \pi}{m}
(j+k+j^{\prime} \! +k^{\prime} \! +j^{\prime \prime} \! 
+k^{\prime \prime} \! -1-2t)(r+s)}
\! \! - \! \! \! \!
\sum^{j+k-1}_{t=0} \! \! e^{\frac{i \pi}{m}
(-j-k-j^{\prime} \! -k^{\prime} \! +j^{\prime \prime} \! 
+k^{\prime \prime} \! +1+2t)(r+s)}
\bigg\}.  \nonumber \\
\label{eqn:summand2}
\end{eqnarray}
}

For $(r,s) \in NS^{(m)}$, we put $p:=r-s$ and $q:=r+s$.
Then $p$ and $q$ are integers satisfying $0 < q < m$, $-q < p < q$ and
$p+q \equiv 1 \,\ \text{mod} \, 2$. So from \eqref{eqn:summand1} and
\eqref{eqn:summand2}, the fusion coefficient 
$N_{(j,k), (j^{\prime}, k^{\prime}), 
(j^{\prime \prime}, k^{\prime \prime})}$
is given in terms of $I(\,\ , \,\ )$ as follows :
\begin{eqnarray}
N_{(j,k), (j^{\prime}, k^{\prime}), 
(j^{\prime \prime}, k^{\prime \prime})}
=\frac{1}{2m^2}(-A_1-A_2+A_3+A_4),
\label{eqn:nuvw}
\end{eqnarray}
where 
\begin{eqnarray*}
A_1 &:=& \sum^{j+k-1}_{t=0}I \! \left(\frac{\pi}{m}
((j \! \! + \! \! j^{\prime} \! \! + \! \! j^{\prime \prime}) \! - \! 
(k \! \! + \! \! k^{\prime} \! \! + \! \! k^{\prime \prime})),
\frac{\pi}{m}(j \! \! + \! \! k \! \! + \! \! j^{\prime} \! \! 
+ \! \! k^{\prime} \! \! + \! \! j^{\prime \prime} \! \! 
+ \! \! k^{\prime \prime} \! \! - \! \! 1 \! \! - \! \! 2t)\right), \cr
A_2 &:=& \sum^{j+k-1}_{t=0}I \! \left(\frac{\pi}{m}
((j \! \! + \! \! j^{\prime} \! \! + \! \! j^{\prime \prime}) \! - \! 
(k \! \! + \! \! k^{\prime} \! \! + \! \! k^{\prime \prime})),
\frac{\pi}{m}(j \! \! + \! \! k \! \! - \! \! j^{\prime} \! \! 
- \! \! k^{\prime} \! \! - \! \! j^{\prime \prime} \! \! 
- \! \! k^{\prime \prime} \! \! - \! \! 1 \! \! - \! \! 2t)\right), \cr
A_3 &:=& \sum^{j+k-1}_{t=0}I \! \left(\frac{\pi}{m}
((j \! \! + \! \! j^{\prime} \! \! + \! \! j^{\prime \prime}) \! - \! 
(k \! \! + \! \! k^{\prime} \! \! + \! \! k^{\prime \prime})),
\frac{\pi}{m}(j \! \! + \! \! k \! \! + \! \! j^{\prime} \! \! 
+ \! \! k^{\prime} \! \! - \! \! j^{\prime \prime} \! \! 
- \! \! k^{\prime \prime} \! \! - \! \! 1 \! \! - \! \! 2t)\right), \cr
A_4 &:=& \sum^{j+k-1}_{t=0}I \! \left(\frac{\pi}{m}
((j \! \! + \! \! j^{\prime} \! \! + \! \! j^{\prime \prime}) \! - \! 
(k \! \! + \! \! k^{\prime} \! \! + \! \! k^{\prime \prime})),
\frac{\pi}{m}(j \! \! + \! \! k \! \! - \! \! j^{\prime} \! \! 
- \! \! k^{\prime} \! \! + \! \! j^{\prime \prime} \! \! 
+ \! \! k^{\prime \prime} \! \! - \! \! 1 \! \! - \! \! 2t)\right).
\end{eqnarray*}

Since $A_1=A_2$ and $A_3=A_4$ by $I(a,b)=I(a,-b)$, the formula 
\eqref{eqn:nuvw} is rewritten as
\begin{equation}
N_{(j,k), (j^{\prime}, k^{\prime}), 
(j^{\prime \prime}, k^{\prime \prime})}=\frac{1}{m^2}(-A_1+A_3).
\label{eqn:nuvw2}
\end{equation}

We note that 
both $j+j^{\prime}+j^{\prime \prime}$ and $k+k^{\prime}+k^{\prime \prime}$
are half of odd integers when $(u, u^{\prime}, u^{\prime \prime}) 
\in \fff^{(m)}$.

In case when $(j+j^{\prime}+j^{\prime \prime})-(k+k^{\prime}
+k^{\prime \prime}) \not\in m \zzz$, all terms in the right-hand sides 
of $A_1$ $\sim$ $A_4$ are equal to $0$ by Lemma 2.2.3), 
giving $N_{(j,k), (j^{\prime}, k^{\prime}), 
(j^{\prime \prime}, k^{\prime \prime})}=0$.
So we need only to consider the case when $(j+j^{\prime}+j^{\prime \prime})
-(k+k^{\prime}+k^{\prime \prime}) \in m \zzz$.  First assume that 
$(j+j^{\prime}+j^{\prime \prime})-(k+k^{\prime}+k^{\prime \prime}) 
\in 2m \zzz$, and look at $A_1$. Then by Lemma 2.2.1) and Lemma 2.2.2), 
one sees that $A_1 = -m(j+k)+m^2$ 
if  either $0$ or $2m$ is contained in the sequence 
$j+k+j^{\prime}+k^{\prime}+j^{\prime \prime}+k^{\prime \prime}-1-2t$ 
($t = 0, \, 1, \, \cdots \, , j+k-1$) namely if 
$-j - k + j^{\prime} + k^{\prime} + j^{\prime \prime} 
+ k^{\prime \prime} < 0$ 
or $j + k + j^{\prime} + k^{\prime} + j^{\prime \prime} 
+ k^{\prime \prime} > 2m$, and $A_1 = -m(j+k)$ otherwise.
When $(j+j^{\prime}+j^{\prime \prime})
-(k+k^{\prime}+k^{\prime \prime}) = \pm m$,
we make use of Lemma 2.2.1') and Lemma 2.2.2').
Applying the same argument to $A_3$, one obtains the following :

\begin{thm}{Lemma 2.3}
\begin{enumerate}
\item[{\rm 1)}] \,\ 
In case when  $(j+j^{\prime}+j^{\prime \prime})-(k+k^{\prime}
+k^{\prime \prime}) \in 2m \zzz$ ;
{\allowdisplaybreaks %
\begin{align*}
&A_1 = 
\begin{cases}
-m(j+k)+m^2 &\,\ \quad \text{if} \,\ 
j^{\prime}+k^{\prime}+
j^{\prime \prime}+k^{\prime \prime} < j+k \\
&\quad \text{or} \,\
j \! + \! k \! + \! j^{\prime} \! + \! k^{\prime} \! +
\! j^{\prime \prime} \! + \! k^{\prime \prime} > 2m, \\
-m(j+k) &\,\ \quad \text{otherwise},
\end{cases}  \\
&A_3 = 
\begin{cases}
-m(j+k)+m^2 &\,\ \quad \text{if} \,\ 
j^{\prime}+k^{\prime}<
j+k+j^{\prime \prime}+k^{\prime \prime} \\
&\! \! \! \! \quad \text{and} \,\
j^{\prime \prime}+k^{\prime \prime}<
j+k+j^{\prime}+k^{\prime}, \\
-m(j+k) &\,\ \quad \text{otherwise}.
\end{cases}
\end{align*}
}
\item[{\rm 2)}] \,\ 
In case when  $(j+j^{\prime}+j^{\prime \prime})-(k+k^{\prime}
+k^{\prime \prime})= \pm m$ ;
{\allowdisplaybreaks %
\begin{align*}
&A_1 = 
\begin{cases}
m(j+k)-m^2 &\,\ \quad \text{if} \,\ 
- \! j \! - \! k \! + \! j^{\prime} \! + \! k^{\prime} \! + \! 
j^{\prime \prime} \! + \! k^{\prime \prime} <m, \\
m(j+k) &\,\ \quad \text{otherwise},
\end{cases}  \\
&A_3 = 
\begin{cases}
m(j+k) &\,\ \quad \text{if} \,\ 
j+k+j^{\prime}+k^{\prime}-
j^{\prime \prime}-k^{\prime \prime}<m \\
& \! \! \! \! \quad \text{and} \,\
j+k-j^{\prime}-k^{\prime}+
j^{\prime \prime}+k^{\prime \prime}<m, \\
m(j+k)-m^2 &\,\ \quad \text{otherwise}.
\end{cases}
\end{align*}
}
\end{enumerate}
\end{thm}

Now Theorem 2.1 follows immediately from \eqref{eqn:nuvw2} and Lemma 2.3.
We remark that the inequality
$$(m-j-k)+(m-j^{\prime}-k^{\prime})
+(m-j^{\prime \prime}-k^{\prime \prime})<2m$$
i.e.,
$$j+k+j^{\prime}+k^{\prime}+j^{\prime \prime}+k^{\prime \prime}>m$$
holds automatically when $(u, u^{\prime}, u^{\prime \prime})$
satisfies the condition (F2).

We define the multiplication on 
the $\zzz$-span $\ffff^{(m)}$ of $A^{(m)}$ by
\begin{eqnarray}
u \cdot u^{\prime} = \sum_{\Sb u^{\prime \prime} \in A^{(m)} \\
\text{s.t.} \,\ (u, u^{\prime}, u^{\prime \prime}) \in \fff^{(m)} \endSb} 
N_{u, u^{\prime}}^{u^{\prime \prime}} u^{\prime \prime}
\label{eqn:fusionalg}
\end{eqnarray}
for $u, u^{\prime} \in A^{(m)}$.  Then the unitarity \eqref{eqn:unitary} 
implies that this multiplication is associative, and so $\ffff^{(m)}$ is 
an associative and commutative $\zzz$-algebra with the 
unit $(\frac{1}{2}, \frac{1}{2})$.

\section*{3. Examples : fusion algebras when m=2, 3 and 4.}

[I] \,\ The case $m=2$ is quite simple since 
$NS^{(2)}= \{(\frac{1}{2}, \frac{1}{2})\}$ and $R^{(2)}= \{(1, 0)\}$.
For simplicity, we put $x_0 := (\frac{1}{2}, \frac{1}{2})$ and 
$x_1:= (1,0)$.  Then $S^{(2)}_{x_0, x_0}=S^{(2)}_{x_0, x_1}=1$ and so 
the fusion algebra $\ffff^{(2)}$ is isomorphic to $\zzz/2\zzz$ with 
the multiplication $x_i \cdot x_j = x_{i+j \, \text{mod} \, 2}$.

\bigskip\noindent
[II] \,\ The case $m=3$;  \,\ In this case, 
$$NS^{(3)} = \left\{\left(\frac{1}{2}, \frac{1}{2}\right),
\left(\frac{1}{2}, \frac{3}{2}\right),
\left(\frac{3}{2}, \frac{1}{2}\right)\right\}  \quad \text{and} \quad
R^{(3)} = \{(1,0),(2,0),(1,1)\}.$$
We put 
{\allowdisplaybreaks %
\begin{align*}
x_0 &:= \left(\frac{1}{2}, \frac{1}{2}\right), \quad 
x_1 := \left(\frac{1}{2}, \frac{3}{2}\right),
\quad x_2 := \left(\frac{3}{2}, \frac{1}{2}\right),  \\
x_3 &:=(1,0), \quad \quad \,\ x_4 :=(2,0), \quad \quad \,\ x_5 := (1,1).
\end{align*}
}
Then the transformation matrix $\left(S^{(3)}_{x_j, x_k}\right)$ is 
given by the following table :
\begin{center}
\begin{tabular}{l||c|c|c|c|c|c}
\noalign{\hrule height0.8pt}
\hfil
& $x_0$ & $x_1$ & $x_2$ & $x_3$ & $x_4$ & $x_5$ \\
\hline
$x_0$ &
$\frac{\sqrt{3}}{3}$ &
$\frac{\sqrt{3}}{3}$ &
$\frac{\sqrt{3}}{3}$ &
$\frac{\sqrt{3}}{3}$ &
$\frac{\sqrt{3}}{3}$ &
$\frac{\sqrt{3}}{3}$   \\
$x_1$ &
$\frac{\sqrt{3}}{3}$ &
$\frac{-\sqrt{3}}{3}\omega$ &
$\frac{\sqrt{3}}{3}\omega^2$ &
$\frac{-\sqrt{3}}{3}\omega^2$ &
$\frac{\sqrt{3}}{3}\omega$ &
$\frac{-\sqrt{3}}{3}$   \\
$x_2$ &
$\frac{\sqrt{3}}{3}$ &
$\frac{\sqrt{3}}{3}\omega^2$ &
$\frac{-\sqrt{3}}{3}\omega$ &
$\frac{\sqrt{3}}{3}\omega$ &
$\frac{-\sqrt{3}}{3}\omega^2$ &
$\frac{-\sqrt{3}}{3}$  \\
\noalign{\hrule height0.8pt}
\end{tabular}
\end{center}
where $\omega := e^{\frac{i \pi}{3}} = \dfrac{1+i \sqrt{3}}{2}$.
From this the fusion algebra $\ffff^{(3)}$ is obtained as follows:
\begin{center}
\begin{tabular}{l||c|c|c|c|c|c}
\noalign{\hrule height0.8pt}
\hfil
& $x_0$ & $x_1$ & $x_2$ & $x_3$ & $x_4$ & $x_5$ \\
\hline
\hline
$x_0$ & $x_0$ & $x_1$ & $x_2$ & $x_3$ & $x_4$ & $x_5$ \\
$x_1$ & $x_1$ & $x_2$ & $x_0$ & $x_5$ & $x_3$ & $x_4$ \\
$x_2$ & $x_2$ & $x_0$ & $x_1$ & $x_4$ & $x_5$ & $x_3$ \\
\hline
$x_3$ & $x_3$ & $x_5$ & $x_4$ & $x_1$ & $x_0$ & $x_2$ \\
$x_4$ & $x_4$ & $x_3$ & $x_5$ & $x_0$ & $x_2$ & $x_1$ \\
$x_5$ & $x_5$ & $x_4$ & $x_3$ & $x_2$ & $x_1$ & $x_0$ \\
\noalign{\hrule height0.8pt}
\end{tabular}
\end{center}

Putting
$$y_0:=x_0, \,\
y_1:=x_3, \,\
y_2:=x_1, \,\
y_3:=x_5, \,\
y_4:=x_2, \,\
y_5:=x_4,$$
one has $y_j \cdot y_k = y_{j+k \, \text{mod} \, 6}$ so that the fusion 
algebra $\ffff^{(3)}$ is isomorphic to the group algebra over the cyclic 
group $\zzz/6\zzz$ of order $6$.

\bigskip\noindent
[III] \,\ The case $m=4$;  \,\ In this case, 
\begin{eqnarray*}
NS^{(4)} &=& \left\{\left(\frac{1}{2}, \frac{1}{2}\right),
\left(\frac{1}{2}, \frac{3}{2}\right),
\left(\frac{3}{2}, \frac{1}{2}\right),
\left(\frac{1}{2}, \frac{5}{2}\right),
\left(\frac{5}{2}, \frac{1}{2}\right),
\left(\frac{3}{2}, \frac{3}{2}\right)\right\} \cr
R^{(4)} &=& \{(1,0),(2,0),(3,0),(1,1),(2,1),(1,2)\}.
\end{eqnarray*}

We put 
{\allowdisplaybreaks %
\begin{align*}
x_0 &:= \left(\frac{1}{2}, \frac{1}{2}\right), \qquad 
x_1 := \left(\frac{1}{2}, \frac{3}{2}\right), \qquad 
x_2 := \left(\frac{3}{2}, \frac{1}{2}\right), \\
x_3 &:= \left(\frac{1}{2}, \frac{5}{2}\right), \qquad
x_4 := \left(\frac{5}{2}, \frac{1}{2}\right), \qquad
x_5 := \left(\frac{3}{2}, \frac{3}{2}\right), \\
x_6 &:=(1,0), \qquad \quad \,\ x_7 :=(2,0), 
\qquad \quad \,\ x_8 :=(3,0), \\
x_9 &:=(1,1), \qquad \quad \,\ x_{10} :=(2,1), \qquad \quad x_{11} :=(1,2).
\end{align*}
}
Then the transformation matrix $\left(S^{(4)}_{x_j, x_k}\right)$ is given 
by the following table :
{\tabcolsep=2.9pt
\begin{center}
\begin{tabular}{l||c|c|c|c|c|c|c|c|c|c|c|c}
\noalign{\hrule height0.8pt}
\hfil
& $x_0$ & $x_1$ & $x_2$ & $x_3$ & $x_4$ & $x_5$ 
& $x_6$ & $x_7$ & $x_8$ & $x_9$ & $x_{10}$ & $x_{11}$ \\
\hline
$x_0$ &
$\frac{\sqrt{2}}{4}$ &
$\frac{1}{2}$ &
$\frac{1}{2}$ &
$\frac{\sqrt{2}}{4}$ &
$\frac{\sqrt{2}}{4}$ &
$\frac{\sqrt{2}}{4}$ &
$\frac{\sqrt{2}}{4}$ &
$\frac{1}{2}$ &
$\frac{\sqrt{2}}{4}$ &
$\frac{1}{2}$ &
$\frac{\sqrt{2}}{4}$ &
$\frac{\sqrt{2}}{4}$ \\
$x_1$ &
$\frac{1}{2}$ &
$0$ &
$0$ &
$\frac{-1}{2}i$ &
$\frac{1}{2}i$ &
$\frac{-1}{2}$ &
$\frac{1}{2}\theta^{-1}$ &
$0$ &
$\frac{1}{2}\theta$ &
$0$ &
$\frac{-1}{2}\theta^{-1}$ &
$\frac{-1}{2}\theta$ \\
$x_2$ &
$\frac{1}{2}$ &
$0$ &
$0$ &
$\frac{1}{2}i$ &
$\frac{-1}{2}i$ &
$\frac{-1}{2}$ &
$\frac{1}{2}\theta$ &
$0$ &
$\frac{1}{2}\theta^{-1}$ &
$0$ &
$\frac{-1}{2}\theta$ &
$\frac{-1}{2}\theta^{-1}$ \\
$x_3$ &
$\frac{\sqrt{2}}{4}$ &
$\frac{-1}{2}i$ &
$\frac{1}{2}i$ &
$\frac{-\sqrt{2}}{4}$ &
$\frac{-\sqrt{2}}{4}$ &
$\frac{\sqrt{2}}{4}$ &
$\frac{-\sqrt{2}}{4}i$ &
$\frac{1}{2}$ &
$\frac{\sqrt{2}}{4}i$ &
$\frac{-1}{2}$ &
$\frac{-\sqrt{2}}{4}i$ &
$\frac{\sqrt{2}}{4}i$ \\
$x_4$ &
$\frac{\sqrt{2}}{4}$ &
$\frac{1}{2}i$ &
$\frac{-1}{2}i$ &
$\frac{-\sqrt{2}}{4}$ &
$\frac{-\sqrt{2}}{4}$ &
$\frac{\sqrt{2}}{4}$ &
$\frac{\sqrt{2}}{4}i$ &
$\frac{1}{2}$ &
$\frac{-\sqrt{2}}{4}i$ &
$\frac{-1}{2}$ &
$\frac{\sqrt{2}}{4}i$ &
$\frac{-\sqrt{2}}{4}i$ \\
$x_5$ &
$\frac{\sqrt{2}}{4}$ &
$\frac{-1}{2}$ &
$\frac{-1}{2}$ &
$\frac{\sqrt{2}}{4}$ &
$\frac{\sqrt{2}}{4}$ &
$\frac{\sqrt{2}}{4}$ &
$\frac{\sqrt{2}}{4}$ &
$\frac{-1}{2}$ &
$\frac{\sqrt{2}}{4}$ &
$\frac{-1}{2}$ &
$\frac{\sqrt{2}}{4}$ &
$\frac{\sqrt{2}}{4}$ \\
\noalign{\hrule height0.8pt}
\end{tabular}
\end{center}
}
\noindent
where $\theta := e^{\frac{i \pi}{4}} = \dfrac{1+i}{\sqrt{2}}$.
From this the fusion algebra $\ffff^{(4)}$ is obtained as follows:
{\tabcolsep=1pt
\begin{center}
\begin{tabular}{l||c|c|c|c|c|c|c|c|c|c|c|c}
\noalign{\hrule height0.8pt}
\hfil
& $x_0$ & $x_1$ & $x_2$ & $x_3$ & $x_4$ & $x_5$ & 
$x_6$ & $x_7$ & $x_8$ & $x_9$ & $x_{10}$ & $x_{11}$ \\
\hline
\hline
$x_0$ & $x_0$ & $x_1$ & $x_2$ & $x_3$ & $x_4$ & $x_5$ &
$x_6$ & $x_7$ & $x_8$ & $x_9$ & $x_{10}$ & $x_{11}$ \\
$x_1$ & $x_1$ & $x_3+x_4$ & $x_0+x_5$ & $x_2$ & $x_2$ & $x_1$ &
$x_9$ & $x_6+x_{10}$ & $x_7$ & $x_8+x_{11}$ & $x_9$ & $x_7$ \\
$x_2$ & $x_2$ & $x_0+x_5$ & $x_3+x_4$ & $x_1$ & $x_1$ & $x_2$ &
$x_7$ & $x_8+x_{11}$ & $x_9$ & $x_6+x_{10}$ & $x_7$ & $x_9$ \\
$x_3$ & $x_3$ & $x_2$ & $x_1$ & $x_5$ & $x_0$ & $x_4$ &
$x_{11}$ & $x_9$ & $x_6$ & $x_7$ & $x_8$ & $x_{10}$ \\
$x_4$ & $x_4$ & $x_2$ & $x_1$ & $x_0$ & $x_5$ & $x_3$ &
$x_8$ & $x_9$ & $x_{10}$ & $x_7$ & $x_{11}$ & $x_6$ \\
$x_5$ & $x_5$ & $x_1$ & $x_2$ & $x_4$ & $x_3$ & $x_0$ &
$x_{10}$ & $x_7$ & $x_{11}$ & $x_9$ & $x_6$ & $x_8$ \\
\hline
$x_6$ & $x_6$ & $x_9$ & $x_7$ & $x_{11}$ & $x_8$ & $x_{10}$ &
$x_3$ & $x_1$ & $x_0$ & $x_2$ & $x_4$ & $x_5$ \\
$x_7$ & $x_7$ & $x_6+x_{10}$ & $x_8+x_{11}$ & $x_9$ & $x_9$ & $x_7$ &
$x_1$ & $x_0+x_5$ & $x_2$ & $x_3+x_4$ & $x_1$ & $x_2$ \\
$x_8$ & $x_8$ & $x_7$ & $x_9$ & $x_6$ & $x_{10}$ & $x_{11}$ &
$x_0$ & $x_2$ & $x_4$ & $x_1$ & $x_5$ & $x_3$ \\
$x_9$ & $x_9$ & $x_8+x_{11}$ & $x_6+x_{10}$ & $x_7$ & $x_7$ & $x_9$ &
$x_2$ & $x_3+x_4$ & $x_1$ & $x_0+x_5$ & $x_2$ & $x_1$ \\
$x_{10}$ & $x_{10}$ & $x_9$ & $x_7$ & $x_8$ & $x_{11}$ & $x_6$ &
$x_4$ & $x_1$ & $x_5$ & $x_2$ & $x_3$ & $x_0$ \\
$x_{11}$ & $x_{11}$ & $x_7$ & $x_9$ & $x_{10}$ & $x_6$ & $x_8$ &
$x_5$ & $x_2$ & $x_3$ & $x_1$ & $x_0$ & $x_4$ \\
\noalign{\hrule height0.8pt}
\end{tabular}
\end{center}
}

One compares this fusion algebra $\ffff^{(4)}$ with that of the affine 
algebra $\widehat{sl}(2, \ccc)$ of level $2$ twisted by the cyclic group 
$\Gamma = (\frac{1}{2}\zzz)/4\zzz$ given in \cite{W}. They may somehow 
resemble each other at a glance, but are quite different.

\end{document}